\documentclass[aps,twocolumn,nofootinbib]{revtex4-1}
\usepackage{amsmath}
\usepackage{amsfonts}
\usepackage{amssymb}
\usepackage{graphicx}
\usepackage{bm}
\usepackage{color}

\begin{document} 

\title{\bf An alternate view of complexity in $k$-SAT problems }

\author{Supriya Krishnamurthy}
\affiliation{Department of Physics, Stockholm University, SE- 106 91, 
Stockholm, Sweden} 
\author{Sumedha}
\affiliation{National Institute of Science Education and Research, Institute of Physics Campus, Bhubaneswar, Orissa- 751 005, India}

\date{\today}

\begin{abstract}
The satisfiability threshold for constraint satisfaction problems is
that value of the ratio of constraints (or clauses) to variables, 
above which the probability that a random instance of the problem 
has a solution is zero in the large system limit. 
Two different approaches to obtaining this threshold 
have been discussed in the literature - using first or second-moment 
methods which give rigorous bounds or using the non-rigorous but powerful 
replica-symmetry breaking (RSB) approach, which gives 
very accurate predictions on random graphs.
In this paper, we lay out a different route to obtaining
this threshold on a Bethe lattice. We need
make no assumptions about the solution-space structure, a 
key assumption in the RSB approach. Despite this, 
our expressions and threshold values 
exactly match the best predictions of the cavity method 
under the $1$-RSB (one-step RSB) hypothesis. 
Our method hence provides alternate interpretations as well as 
motivations for the key equations in the RSB approach.
\end{abstract}

\maketitle

\begin{figure}[ht]
  \begin{center} 
  \includegraphics[scale=0.5]{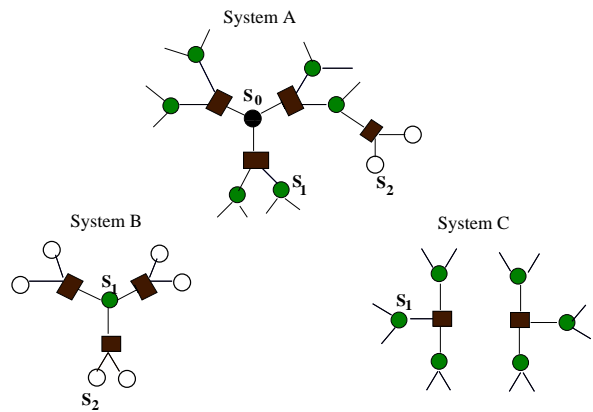}
  \caption{
   The three systems considered in the text are shown. For the example above, $k=3$ and $d+1=3$. In this case, system B is a set of $4$ independent trees of the type shown. Systems A and B differ by exactly  $k=3$ nodes and systems A and C differ by $1$ node.
    \label{fig:model} 
  }
  \end{center}
\end{figure}

The random $k$-satisfiability problem may be stated as follows.
Fix $k$ and consider $N$ boolean variables $x_1,x_2, \cdots, x_N$.
An expression of the sort  $\phi_1 = x_1 \lor x_5 \lor \bar{x_{9}}$ 
is called a clause ($k=3$ in the above example) and evaluates to TRUE, 
if at least one of the $k$ variables evaluates to TRUE, {\i.e.} 
$x_1=1$ or $x_5=1$ or $\bar{x_{9}}=0$. For each clause $i$, the
boolean information set $\eta_1^{i}$, $\eta_2^{i},\cdots,\eta_k^{i}$, 
called a literal assignment, carries information 
on which of the variables in $i$ are negated.
In the most standard version of the $k$-SAT problem, 
the random $k$-SAT, each literal takes the value $0$ or $1$ with
equal probability and hence all literal assignments are equiprobable.
A formula is an expression $\phi_1 \land \phi_2 \cdots \phi_M$ with  
$M$ clauses generated randomly from the $N$ variables. 
This formula has a solution if there is 
any assignment of variables
for which {\em all} $M$ clauses are satisfied or evaluate to TRUE. 
A formula may be represented as a factor graph, with $N$ vertices (or variables), $M$ clauses
(denoted by a filled square in Fig. \ref{fig:model}) 
and (two types of) edges connecting 
the vertices to clauses. If the further constraint is imposed of
each vertex participating only in $d+1$ clauses, then the factor 
graph represents a random regular graph.


Random $k$-satisfiability is only one of many so-called constraint 
satisfaction problems. Several variations have been considered in which 
the constraints (or clauses) have  different forms.
Literal assignments may also be chosen differently.
For example, in the $k$-NAE-SAT problem,
a clause is unsatisfied  both if all variables that participate in it 
evaluate to FALSE, or if they evaluate to TRUE, 
with all literal assignments being equiprobable.

One can vary a formula by varying only the literal assignment 
(or instance) and ask what is the probability that
a randomly chosen literal assignment has a solution. 
If an instance has a solution then it is a 
satisfiable instance.
As the  constraint density ($\alpha = M/N$) increases, this
probability decreases. In the limit of $M,N \rightarrow \infty$, 
the system is known to have a sharp threshold $\alpha_s$ below 
which the probability of finding   satisfiable instances approaches 
$1$ and above which it vanishes\cite{msl,kirkpatrick}.   

The location of $\alpha_s$ has long been an area  
of active research in the computer science field, and is known rigorously    
for $k=2$ \cite{chvataletal}. For higher $k$, several rigorous results  
on upper and lower bounds to the solvability  
threshold exist \cite{achlioptas_bounds},    
usually obtained using first moment arguments or the 
second moment method \cite{achlioptas1,achlioptas_peres} 
respectively. 

Physicists meanwhile, have tackled this problem using other methods, namely
the replica and cavity approaches from spin-glass theory \cite{mezard_book}.
In these approaches, it is assumed that, for any typical literal assignment, 
the solutions
start to {\it cluster} beyond a certain clustering threshold $\alpha_d$, 
which lies strictly below the satisfiability transition $\alpha_s$.
$\alpha_d$ arises naturally as the point where the cavity recursions
develop a new non-trivial solution.
$\alpha_s$  on the 
other hand, is the location at which the log of the 
{\it number} of solution clusters goes to zero. 
This latter quantity evaluated per variable, 
called {\it complexity}, results in the $1$-RSB prediction
for $\alpha_s$ \cite{mezard-science,mmz}. 
The cavity approach also predicts a number of other thresholds
between the clustering and satisfiability thresholds, such
as the condensation threshold \cite{PNAS} and the
freezing threshold \cite{semerjian}.

These two different theoretical routes for obtaining $\alpha_s$, have 
come together in recent work, with both the existence of clusters 
proven rigorously \cite{achlioptas-coja} as well as the 
$1$-RSB prediction for $\alpha_s$ etablished explicitly 
for the  $k$-NAE-SAT problem \cite{sly} and 
the $k$-SAT problem \cite{coja-oghlan, sly1} for large $k$.
Both approaches obtain $\alpha_s$
by looking at properties of solution clusters.
 In this paper, we obtain instead the {\it same} expression for $\alpha_s$ 
by calculating directly the
fraction of literal assignments 
that have solutions.
We demonstrate how to do this for both $k$-SAT as well as
$k$-NAE-SAT, building on previous work \cite{ss,sss}, where we computed
the fraction of satisfiable literal assignments exactly on trees.
We outline a procedure, equivalent to performing this calculation now
on a {\it Bethe lattice}, which gives us an easy way to derive an analog of the complexity for this whole class of problems.

Our method can easily be applied even to obtaining an expression for the 
moments of the number of solutions on the Bethe lattice. Applying 
the procedure to the first moment results in the replica-symmetric
expression for the complexity. Applying this method to the second moment
results, for the first time to our knowledge, 
in an analog of the complexity (or a rate function) for the second moment.
This expression, derived for a Bethe lattice, again matches
its counterpart on a random graph \cite{achlioptas1} in 
so far as all quantitative 
predictions go, and in addition, brings to light different transitions 
connected with the change in the nature of the overlap of solutions.

{\it \bf The Analog of Complexity} :
We consider first a factor graph which is a 
rooted tree with all nodes having a degree $d+1$.
The boundaries, or leaf-nodes,
are assigned a fixed value $0$ or $1$ randomly, and have a degree $=1$.
Every node on this tree (except the boundary nodes) is the root of its subtree
and considered as such, we can vary the literal assignment of the
edges on this subtree and calculate 
the fraction of instances $P_0$ in which this node can take no value 
(since either value would violate a clause that it participates in), only 
one value ($0$ or $1$) $P_1$ or both values $P_2$ \cite{ss,sss}.
Clearly $P_0 + P_1 + P_2 = 1$. 
The quantity $\sum \log (1 - P_0)$ then, where the sum is over {\it every}
node in the tree, is the log of the probability that an arbitrary assignment of
literals over the whole tree, is satisfiable \cite{ss}. 
In general, for a tree, $P_0$ will depend on the {\it level}
of a node (its distance from the leaves, with the leaves at level $0$).
For an infinite tree, the quantity $P_0$ eventually becomes independent of the level of the node (except for the central node which has $d+1$ subtrees)
and the value it 
takes is given by the fixed point of the tree recursions \cite{ss,sss}.
These tree-recursions are written for the quantity
$Q$; $Q \equiv \frac{P_1}{2(1-P_0)}$ is the fraction of instances (taken 
{\it only} over all satisfiable instances on the sub-tree) 
in which the root can only take
the one value {\it not} satisfying
the clause connecting it to the node at the higher level, 
for {\it any} literal assignment. 
For  $d< d_c$, the fixed-point equation (FPE) only has 
a trivial solution $Q=0$. Above $d_c$ a second
non-trivial stable solution exists at a non-zero value of $Q$.

For an infinite tree, let us now consider
only 'interior' nodes with high enough levels such that
the tree recursions have reached a fixed point. 
For this system, we can use the relation $\alpha = \frac{d+1}{k}$.
$\alpha_d= \frac{d_c+1}{k}$ then indicates the branching beyond which the 
fraction of literal assignments that have solutions goes to $0$.
The value $\alpha_d$ is exactly the same as obtained earlier \cite{mmz}
for the clustering transition
(for reasons we explain a little later). However for our model, 
$\alpha_d$ signifies a satisfiability threshold on the tree.

If however, instead of a tree, we consider a graph which is only 
locally tree-like,
with all nodes having a degree $d+1$ and with neither a {\it central} node nor
a surface, then as detailed below,  this system displays a non-trivial satisfiability threshold exactly as predicted by the cavity method.
This graph is what we mean by a {\it Bethe} lattice.

The problem now is to
calculate the analog of $\sum \log(1-P_0)$ on our Bethe lattice.
We do not expect it to be possible to calculate
$P_0$ node-by-node as we did for the tree. However, if it is possible 
to calculate the  fraction of satisfiable instances for two systems (obeying all the same constraints) differing only by 
a known number of nodes, then a logarithm of the ratio of the two quantities 
can provide an estimate per node.
A general and simple prescription for calculating partition 
functions in this manner on the Bethe lattice has been given by 
Gujrati \cite{gujrati}.
The idea is to consider two separate trees which differ only by a certain number of internal nodes 
but are so constructed that they have exactly the same
number of leaves. The fraction of satisfiable instances is 
exactly calculable 
for both of these systems. A logarithm of the ratio then provides an estimate of
the logarithm of the fraction of satisfiable instances per node. 

The  idea outlined above applied to $k$-SAT  leads to the 
consideration of the following systems (see Fig. \ref{fig:model}).
System A has a central node (or root node)  $s_0$, with $(k-1)(d+1)$ neighbours $s_1$.
Each of these $s_1$ neighbours have $(k-1)d$ other neighbours besides the root 
($s_2$ nodes). System B is a collection of $(k-1)d$ $s_1$ nodes, 
each however with $(k-1)(d+1)$ $s_2$ neighbours. It is easy to see that the two 
situations have the same number of leaf nodes, and differ from each other only 
by $k$ internal nodes ($k-1$ $s_1$ nodes + $1$ $s_0$ node), or $d+1$ clauses.
Alternatively, we could also consider a system C (see Fig. \ref{fig:model}), which 
differs from system A by exactly one node. 
Note that all three systems satisfy the same constraints, namely
all nodes have a degree $d+1$ and all clauses have a degree $k$.

A logarithm of the ratio of the fraction of satisfiable instances of system A to system B (or system C) should then result in a value of this quantity for $k$ nodes (respectively one node). As we will see, this is the quantity that plays the role of complexity.

In terms of the quantity $Q$, the logarithm of the ratio of the two
probabilities (we call this $\Sigma$ in analogy with all the earlier work) is 
\begin{eqnarray}
k\Sigma &=& \log \left\{\frac{f_{d+1}(Q) f_d(Q)^{(k-1)(d+1)}}{f_{d+1}(Q)^{(k-1)(d)}} \right\}
\label{eq:comp_def}
\end{eqnarray}
where $f_{d+1}$ is the probability $1-P_0$ calculated for a node with degree $d+1$ and $f_{d}$ is the equivalent quantity calculated for a node with degree $d$. In what follows, we present the details of our calculations for both $k$-NAE-SAT and $k$-SAT.

{\it \bf $k$-NAE-SAT on the Bethe lattice}:
The recursions for $P_0$, $P_1$ (with $P_2=1-P_0-P_1$) on a rooted tree with branching number $d$ are:
\begin{eqnarray}
P_0 &=& 1+ (1-2Q^{k-1})^{d} - 2 (1-Q^{k-1})^{d} \equiv 1-f_d(Q) \nonumber\\
P_1 &=& 2 (1-Q^{k-1})^{d} - 2 (1-2Q^{k-1})^{d} \nonumber \\
\label{eq:nae}
\end{eqnarray}


The two equations above may be written as a recursion for one single quantity $Q$. The FPE is then
\begin{equation}
Q = \frac{(1-Q^{k-1})^{d} - (1-2Q^{k-1})^{d}}{2(1-Q^{k-1})^{d} - (1-2Q^{k-1})^{d}}
\label{eq:rec_nae}
\end{equation}
Eq. \ref{eq:rec_nae}, with a change of variables, is the same equation derived
in \cite{dall'asta} by the cavity method. Above a critical value  $d_c(k)$,
Eq. \ref{eq:rec_nae} has a second non-trivial solution for any $k > 2$ in which $Q$ is non-zero. 



\begin{figure}[ht]
  \begin{center} 
  \includegraphics[scale=0.5]{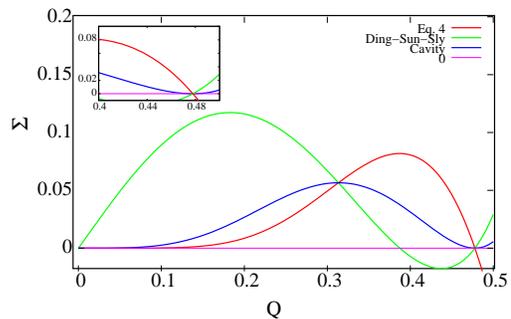}
  \caption{
Eq. \ref{eq:comp_nae} is plotted along with the cavity prediction (from \cite{dall'asta}) as well as the corresponding expression from the  work of Ding {\it et al} \cite{sly}. These three functions are converted
into exactly the same function by  requiring Eq. \ref{eq:rec_nae} to be 
satisfied.
Here all three functions evaluate to $0$ at the non-trivial fixed point value of $Q$ for $d_s=128.6$ for $k=6$, 
  resulting in the prediction for the satisfiability threshold for $6$-NAE-SAT to be $\alpha_s = \frac{d_s+1}{k}= 21.58$. The inset shows a close up of the three functions in the region of interest.
    \label{fig:comp_nae} 
  }
  \end{center}
\end{figure}

For both systems A and B, 
it is easy to calculate the fraction of satisfiable instances exactly and so 
also the logarithm of their ratio (Eq. \ref{eq:comp_def}).
For $k$-NAE-SAT, substituting the expressions from Eq. \ref{eq:nae} 
into Eq. \ref{eq:comp_def}, we get
\begin{eqnarray}
k\Sigma &=& (d+1) \log(1-Q^{k-1})+(d+1)(k-1)\log(2-g(Q)^{d}) \nonumber \\
&+&  (d+1-dk)\log(2-g(Q)^{d+1}) 
\label{eq:comp_nae}
\end{eqnarray}
where $g(Q)= \frac{1-2Q^{k-1}}{1-Q^{k-1}}$. $\Sigma$ is evaluated at the
fixed point of the recursion for $Q$ (Eq. \ref{eq:rec_nae}) 
(see Fig. \ref{fig:comp_nae}). It is easy to show that 
Eq. \ref{eq:comp_nae}  is exactly the same as the $m=0$, $1$-RSB expression for 
complexity, obtained for $k$-NAE-SAT on random regular graphs
by Dall'asta {\it et al} \cite{dall'asta}, 
as well as the exact expression for the 
rate function obtained by Ding {\it et al} \cite{sly} 
for $k$-NAE-SAT (for large $k$) on random regular graphs. 
(Fig. \ref{fig:comp_nae}).

It is interesting at this point to consider the interpretation of
the {\it complexity} function in our case.
When $Q=0$, $\Sigma=0$ and the fraction of satisfiable 
instances is $=1$ for both the numerator and the denominator.
As $d$ increases, beyond a critical value  $d_c$ as explained above, 
Eq. \ref{eq:rec_nae} has two stable solutions, in one of which
$Q$ is non-zero. 
The value of $d_c$ is the same for systems A and B, but
the value of $\Sigma$ is non-zero and positive upto a value
$d_s > d_c$, for non-zero $Q$ (see Fig. \ref{fig:comp_nae}). 
A positive value of $\Sigma$ 
is however not consistent with its interpretation
as the logarithm of a probability, since
in this case, it should either be $=0$ or negative.
For $d<d_c$, $Q=0$ and $\Sigma=0$, which is consistent. 
For $d>d_c$, $\Sigma$ can take a physically acceptable value only
if the chosen solution continues to be $Q=0$, till the value of 
$d$ when $\Sigma$ crosses over to negative values.
This value of $d$ is indeed the 
satisfiability threshold $d_s$, since 
as soon as $Q$ becomes non-zero, the contribution of each node to $\Sigma$ is
negative and the fraction of satisfiable instances
goes to $0$ in the large $N$ limit.



{\it \bf $k$- SAT on the Bethe lattice}:
For completeness we show the results for $k$-SAT as well.
The FPE for the quantity $Q$ for $k$-SAT on a tree with 
branching number $d$ 
is \cite{ss}

\begin{eqnarray}
Q &=& \frac{(1-0.5Q^{k-1})^d - (1-Q^{k-1})^d}{2(1-0.5Q^{k-1})^d - (1-Q^{k-1})^d} 
\label{eq:rec_ksat}
\end{eqnarray}

The function $ f_d(Q) = {2(1-0.5Q^{k-1})^d - (1-Q^{k-1})^d}$,
and substituting this in Eq. \ref{eq:comp_def} we get,
\begin{eqnarray}
k\Sigma &=& (d+1) \left[ \log(1-0.5 Q^{k-1}) + (k-1)\log(2-g(Q)^{d}) \right] \nonumber \\ 
         &+& (d+1-dk)\log(2-g(Q)^{d+1}) 
\label{eq:comp_ksat}
\end{eqnarray}
where $Q$ satisfies the FPE Eq. \ref{eq:rec_ksat} and
$ g(Q)= \frac{1-Q^{k-1}}{1-0.5 Q^{k-1}}$.
The cavity approach predicts for this problem the expression \cite{note}

\begin{eqnarray}
\Sigma^{\prime} &=& \log(2-g(Q)^{d+1}) + (d+1) \log(1-0.5 Q^{k-1}) \nonumber \\
&-& (d+1)(1-1/k)\log(1-Q^{k})   
\label{eq:comp_mmz}
\end{eqnarray}

Again, $\Sigma= \Sigma^{\prime}$ as long as 
the FPE Eq. \ref{eq:rec_ksat} is satisfied.

The second moment method applied to clusters of $k$-SAT solutions
has been used recently to obtain the exact threshold in a regular symmetrized $k$-SAT problem \cite{coja-oghlan} as well as for $k$-SAT on random graphs \cite{sly1}. Both these works confirm the $1$-RSB prediction for the complexity
for these problems, implying in turn that our procedure
on the Bethe lattice gives results which coincide with the exact expressions for random graphs.


{\it \bf First and Second Moment of the total number of solutions}: 
The  procedure detailed above  may be utilized to obtain a 
rate function for any quantity that 
varies exponentially with the number of variables $N$.
We now define our quantity of interest to be the  
first moment $\langle Z \rangle $  and second moment $\langle Z^{2} \rangle $ of the number of solutions, where $Z$ is the total number of solutions for a given literal assignment. 
For the $k$-SAT problem on a tree with branching number 
$d$, fixed boundary nodes and a given literal assignment,  it 
is easy to write a recursion relation for $Z$ as a function of the 
level \cite{ss}. The corresponding 
recursion relation for $k$-NAE-SAT  is only slightly different \cite{lp}.
The first moment $\langle Z \rangle$ is readily obtained from the recursion
and the corresponding rate function results in
the replica-symmetric expression for the
complexity (as expected).




If we follow the same procedure to now obtain a rate function for the
second moment $\langle Z^{2} \rangle$, we obtain the expression
\begin{eqnarray}
\frac{log \langle Z^{2}\rangle}{N} &=& \log(2) + \frac{k-1}{k}(d+1) \log\left( \frac{h^{d}}{2} \right) \nonumber \\
&+& \left(\frac{d+1}{k} -d \right) \log\left[ h^{d+1} + \hat{h}^{d+1} \right]
\label{eq:secmom}
\end{eqnarray}
where the functions $h= \left[(2^{k-1}-1)(1+r)^{k-1} + 0.5 \right]$, 
$\hat{h}= h-0.5$ and $r$ is the solution of  the fixed point
equation (\cite{ss}) 
\begin{equation}
r= \frac{\hat{h}^{d}}{h^{d}}
\label{eq:secmom_fp}
\end{equation}

\begin{figure}[ht]
  \begin{center} 
  \includegraphics[scale=0.5]{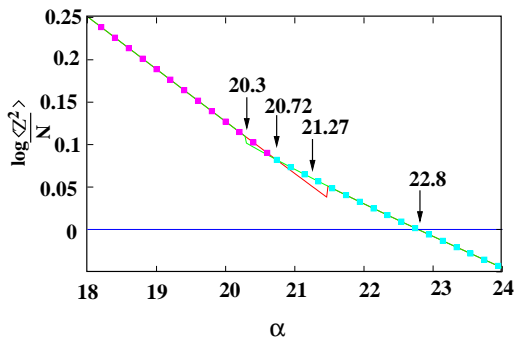}
  \caption{
  The expression for the logarithm of the second moment (per variable) from
  Eq. \ref{eq:secmom} 
is plotted for $k=5$. The various {\it  transitions} are also indicated in the figure with $\alpha = \frac{d+1}{k}$.  The clustering transition, at $ \alpha_d =16.16$ is not shown. At  $ \alpha_f \sim 20.3$, the fixed point equation for $r$ (Eq. \ref{eq:secmom_fp}) develops
a second solution, which is however chosen only at  $ \alpha_0 = 20.72$ where 
the expression for the second moment (Eq.  \ref{eq:secmom}) develops a discontinuity in the
derivative. The satisfiability threshold occurs at  $ \alpha_s = 21.27$.
At $ \alpha_M = 22.8$, the logarithm goes to $0$. Beyond this value, the second moment vanishes in the large $N$ limit.
    \label{fig:sm} 
  }
  \end{center}
\end{figure}

For $k<5$, Eq. \ref{eq:secmom_fp} has only one solution for any $d$.
For $d > d_f$ and for $k>5$, the 
fixed point equation 
develops two solutions (interestingly, as $d$ continues to increase, the
fixed point equation goes back to having only one solution, but this value of $d$ lies well beyond the point where Eq. \ref{eq:secmom} vanishes, for large $k$).The value of $r$ which leads to 
the larger value of the second moment (Eq. \ref{eq:secmom} ) is the chosen one.
At $d = d_o(k)$, this value changes,
leading to a discontinuity in the derivative of the second moment
with respect to $d$ (or $\alpha$); see Fig.\ref{fig:sm}.

$d_o$ occurs before the satisfiability threshold for $k>5$, though
well after the clustering threshold.  In terms of the overlap between two solutions, the discontinuity of the derivative of the second moment
translates to  a discontinous transition from a low to a high value, at $d_o$.
This was first remarked in \cite{troyansky}, and is also exhibited by the
exact expression of the second moment in terms of the overlap, 
for a random graph \cite{achlioptas1}.
The precise value $\alpha_M$ at
which Eq. \ref{eq:secmom} equals zero, also matches the random-graph values 
for all $k$. It would be very interesting to understand if $d_f$ and $d_o$
are connected to the condensation \cite{PNAS} and freezing \cite{semerjian} 
transitions.
We include a table of values of $\alpha$ for the different transitions 
for varying $k$ (see table \ref{tab:a}).





\begin{table}
\caption{The different transitions we obtain in $k$-SAT. $\alpha_d$ and $\alpha_s$ are the clustering and satisfiability thresholds respectively and match the cavity predictions. $\alpha_f$ is the value at which Eq. \ref{eq:secmom_fp} develops a second solution and $\alpha_o$ is the value at which this solution is chosen. $\alpha_M$ is the point after which the second moment vanishes.}

\label{tab:a}
	\centering
		\begin{tabular}{|l|l|l|l|l|l|}  
\hline
		$K$ &  $\alpha_d$ & $\alpha_f$ & $\alpha_s $ & $\alpha_o$ & $\alpha_M$ \\
		\hline  $2$ & $1.5$ & $-$ & $1.5$ & $-$ & $2.8$ \\ \hline  $3$ &
		$4.16$ & $-$ & $4.55$ & $-$ & $5.83$ \\ \hline $4$ & $8.4$ & $-$ & $10.15$ &$-$& $11.56$  \\
		\hline $5$ & $16.2$ & $20.3 $  & $21.26$ & $20.7$ & $22.8$   \\ \hline $6$ &
		$30.5$ & $38.5$ & $43.41$ & $42.95$ & $45$  \\ \hline $7$ & $57.28$ & $72.1$ & $87.84$ & $87.35$ & $89.42$   \\ \hline $8$ & $107.13$ & $134.2$ & $176.57$ & $176.06$ & $178.1$   \\
		\hline
		\end{tabular}
\end{table}

In analogy with the work on tree reconstruction \cite{mezard-montanari}, which related $\alpha_d$  to a process on trees, we are able to give an interpretation
for both $\alpha_d$ and $\alpha_s$ entirely in terms of the fraction of
satisfiable instances. The calculations though
explicitly performed on a Bethe lattice, match in every case, the relevant
expressions on random (regular) graphs. 
Our methods are easily applicable even to $k$-SAT problems on random graphs with
other degree distributions, or for models with variable clause sizes, making this perhaps a useful tool for studying real-world SAT applications.
{\it \bf Acknowledgements }: We would like to thank
Deepak Dhar and Guilhem Semerjian for a critical reading of the manuscript and Cris Moore for several helpful suggestions.





\begin{thebibliography}{99}
\bibitem{note} We have not seen this expression explicitly written anywhere, but
it follows from an application of the methods of \cite{dall'asta}.
\end{thebibliography}
\end{document}